\documentclass[manuscript]{acmart}

\usepackage{subcaption}
\usepackage{caption}
\usepackage{graphicx}
\AtBeginDocument{%
  }

\setcopyright{acmlicensed}
\copyrightyear{2024}
\acmYear{2024}
\acmDOI{XXXXXXX.XXXXXXX}

\acmConference[Conference acronym 'XX]{Make sure to enter the correct
  conference title from your rights confirmation emai}{June 03--05,
  2018}{Woodstock, NY}

\begin{document}

\title{AuditNet: A Conversational AI-based Security Assistant [DEMO]}


\author{Shohreh Deldari}
\email{s.deldari@unsw.edu.au}
\affiliation{%
  \institution{University of New South Wales (UNSW) and CSCRC, Australia}
}

\author{Mohammad Goudarzi}
\email{m.goudarzi@unsw.edu.au}
\affiliation{%
  \institution{University of New South Wales (UNSW) and CSCRC, Australia}
}

\author{Aditya Joshi}
\email{aditya.joshi@unsw.edu.au}
\affiliation{%
  \institution{University of New South Wales (UNSW) and CSCRC, Australia}
}

\author{Arash Shaghaghi}
\email{a.shaghaghi@unsw.edu.au}
\affiliation{%
  \institution{University of New South Wales (UNSW) and CSCRC, Australia}
}

\author{Simon Finn}
\email{sifinn@cisco.com}
\affiliation{%
  \institution{Cisco}
}

\author{Flora D. Salim}
\email{flora.salim@unsw.edu.au}
\affiliation{%
  \institution{University of New South Wales (UNSW) and CSCRC, Australia}
}

\author{Sanjay Jha}
\email{sanjay.jha@unsw.edu.au}
\affiliation{%
  \institution{University of New South Wales (UNSW) and CSCRC, Australia}
}

\renewcommand{\shortauthors}{Deldari .et al.}


\begin{abstract}
In the age of information overload, professionals across various fields face the challenge of navigating vast amounts of documentation and ever-evolving standards. Ensuring compliance with standards, regulations, and contractual obligations is a critical yet complex task across various professional fields. We propose a versatile conversational AI assistant framework designed to facilitate compliance checking on the go, in diverse domains, including but not limited to network infrastructure, legal contracts, educational standards, environmental regulations, and government policies. By leveraging retrieval-augmented generation using large language models, our framework automates the review, indexing, and retrieval of relevant, context-aware information, streamlining the process of verifying adherence to established guidelines and requirements. This AI assistant not only reduces the manual effort involved in compliance checks but also enhances accuracy and efficiency, supporting professionals in maintaining high standards of practice and ensuring regulatory compliance in their respective fields.
We propose and demonstrate AuditNet, the first conversational AI security assistant designed to assist IoT network security experts by providing instant access to security standards, policies, and regulations. 
\end{abstract}





\received{24 June 2024}

\maketitle

\section{Introduction}

The traditional methods of managing and interpreting extensive documentation for compliance purposes are labor-intensive, time-consuming, and require specialized expertise. This process is typically performed manually, involving meticulous reading and indexing of each reference document. Additionally, because users need to consider multiple references to verify compliance, cross-referencing between documents becomes essential. The evolving nature of the security landscape demands continuous monitoring and adaptation to new standards and regulations. Consequently, these manual methods are not scalable to cover numerous documents and cannot easily adapt to the constant changes and new regulations. Interpreting user queries and cross-referencing them to various references is also a challenging and time-consuming task for domain experts. Similarly, for network professionals who interact with client-facing projects, it may not be feasible to refer to specific standards and check for compliance with detailed network configuration files. Providing them with an easy, versatile assistant that is accessible via a mobile phone 
to streamline the process.

To address these challenges, we introduce a versatile conversational AI assistant framework designed to streamline compliance checking across various fields. Although our focus in this work is on cybersecurity compliance checks, the proposed framework can be transferred to a wide range of domains, such as legal contracts, educational standards, environmental regulations, and government policies.
In this work, we consider cybersecurity compliance checks in telecommunication and inter-networking infrastructure as the motivational application of our framework. Therefore, the final users of the system can be Security experts, security auditors or network designers who need to audit and check whether the existing infrastructure is compliant with the policies and regulations established by the company or cyber-security standard organizations. Considering the fact that the field of network security is continuously evolving, with frequent updates to standards and regulations requiring professionals to stay current to ensure compliance and protect their organizations. 
Traditional manual methods of cross-referencing and interpreting security standards are labour-intensive, may not be scalable, and require specialized expertise to keep up with rapid changes and new regulations. To address these challenges, we propose an automated conversational AI system integrated with machine learning capabilities to simplify the comprehension and application of complex network security standards. The proposed system, named \textit{AuditNet}, includes the following features:

\begin{itemize}
    \item Automated Processing: Automates the review and indexing of standard documents, dynamically building a knowledge base from these texts.
    \item Adaptability and Scalability: Efficiently adapts to new regulations and scales to accommodate various security standards and stakeholder needs.
    \item User Interaction: Interacts directly with end-users, processing their queries and providing accurate responses based on the existing knowledge base.
\end{itemize}

AuditNet aims to enhance compliance and streamline the process of staying updated with network security regulations, ultimately supporting network professionals in maintaining robust security practices. This framework represents a significant advancement in the efficient and accurate handling of compliance-related tasks, providing tailored, actionable insights to ensure regulatory adherence across multiple domains. To achieve this, we leverage existing open-source Large Language Models (LLM) to understand users' intentions and relate them to suitable rules/policies/requirements requested in the large corpus of guidelines. Next, we benefit again from the capability of the LLM to generate an accurate answer and also bridge the gap between the policies and the underlying infrastructure by pointing out which parts of the infrastructure are affected by this policy. Hence the generated response effectively directs the security auditor to the relevant standard implications (policy) and the infrastructure implementation at once.

In the following sections, we first introduce several related techniques used by the community and then cover the background of the models and tools employed in our framework. Next, we present our framework, including the data and our data augmentation approach, and finally, we evaluate our technique. This structure provides a comprehensive overview of the context, methodology, and effectiveness of our approach.

\section{Background and Related Works}
In recent years, Large Language Models (LLMs) have revolutionized the field of natural language processing, showcasing exceptional performance across a diverse array of applications and domains such as medical \cite{guo2023improving}, recruitment \cite{recruiter}, smart-assistants \cite{zhang2024notellm}, network infrastructure, and cyber-security \cite{garza2023assessing,park2023pretrained}. These general models are powered by vast amounts of generic data and extensive pre-training. 
However, when it comes to applications with limited access to high-quality labeled data or specific domain applications with sparse data, the primary challenge is effectively leveraging these general models. LLMs are highly capable tools for generating realistic text samples based on existing data. For NLP tasks, generating data with LLMs can include paraphrasing text, creating alternative question-answer pairs, or generating new sentences. Producing diverse representations of input data allows models to learn various ways to express the same underlying concepts, thereby increasing their adaptability to real-world data variations. We utilize LLMs in both data augmentation and inference steps to develop our system, despite having limited access to data.

\begin{figure}
    \centering
    \includegraphics[height=2.5in]{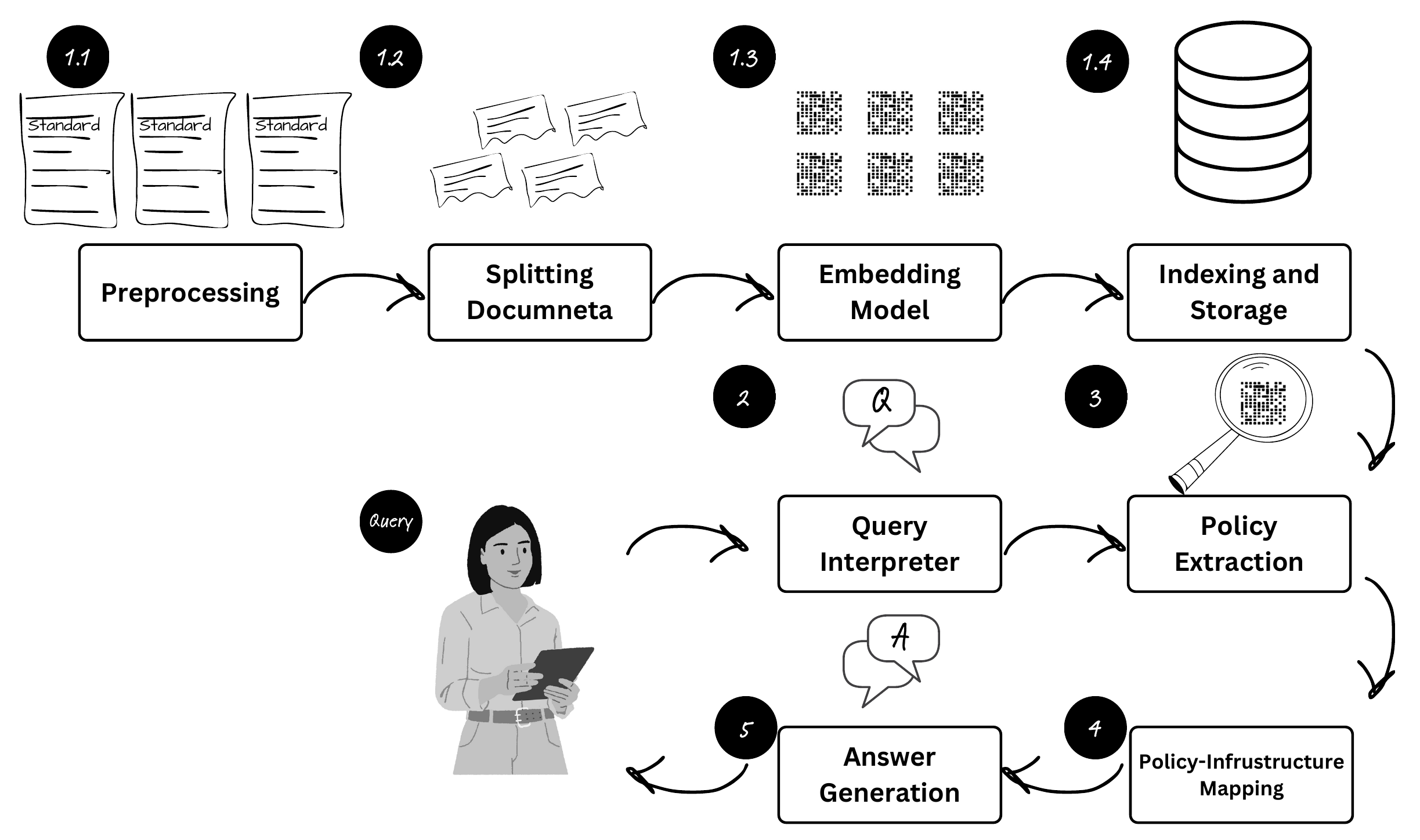}
    \caption{AuditNet Framework}
    \label{fig:pipeline}
\end{figure}

\section{AuditNet Framework}

In this section, we first outline the manual process currently used for compliance checks on IoT network infrastructure. We then introduce our model's pipeline.

In the manual process, security experts must thoroughly understand relevant security standard documents to grasp the structure and requirements needed for implementation and design. This involves extensive note-taking and documentation to track compliance details. Each security question or compliance check raised by users requires reviewing numerous security standards and compiled indexes to extract information like policy definitions, implementation specifics, and required protocols. Experts then compile a list of contextual information from the infrastructure to verify adherence to these protocols.
In our work, we automate the entire process: processing security standards and regulations, interpreting user queries, and aligning standard rules and policies with IoT infrastructure implementation in AuditNet. The overall framework of AuditNet (illustrated in Figure \ref{fig:pipeline}) is explained below:

\subsection{Processing Standard Documents}
The process involves splitting the documents into smaller chunks of information, embedding them in vector format using a pre-trained embedding model and then storing them in an efficiently accessible database (Vector store). 

\subsubsection{Preprocessing.} PDF, although a common document format, may not be optimal for structured content interpretation. Consequently, we devised a parser specifically tailored for reading and converting PDF text into Markdown format. To facilitate structured content interpretation, we developed a custom parser that employs regular expressions to extract relevant components such as section titles and body text.

\subsubsection{Splitting documents:} 
\begin{figure}
\centering
\begin{minipage}{.29\textwidth}
  \centering
  \includegraphics[width=.98\linewidth]{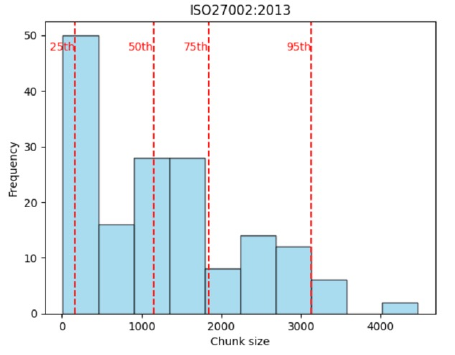}
  \captionof{figure}{Block of texts in document A.}
  \label{fig:iso_split}
\end{minipage}%
\begin{minipage}{.29\textwidth}
  \centering
  \includegraphics[width=.9\linewidth]{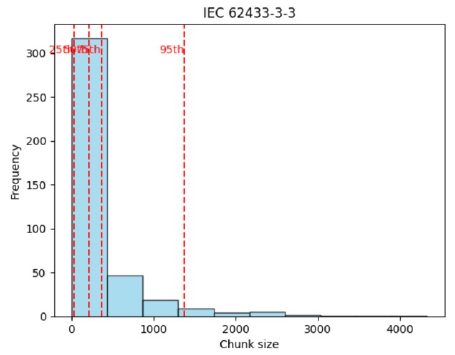}
  \captionof{figure}{Block of texts in Standard B.}
  \label{fig:iec_split}
\end{minipage}%
\begin{minipage}{.4\textwidth}
  \centering
  \includegraphics[width=.9\linewidth]{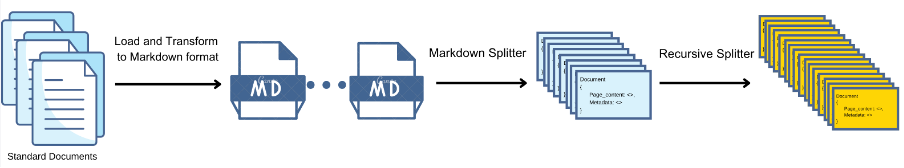}
  \captionof{figure}{Document splitting process}
  \label{fig:splitting}
\end{minipage}
\end{figure}

We use Markdown features to retain the document's structure by splitting the text into smaller blocks, each corresponding to a single subsection. The length of these blocks, which are the smallest divisions, can vary significantly, ranging from a few lines to several pages. The distribution of the lengths of text blocks in two sample standard documents is shown in Figures \ref{fig:iso_split} and \ref{fig:iec_split}. This variation can cause issues such as misalignment with the context size required by language models and embedding tools, or it can degrade performance. Ideally, each split should not only keep semantically related text together but also be independently meaningful without relying on subsequent splits.
We initially divide the text based on the document’s structural divisions (i.e., by sections) to ensure semantic coherence. To address the issue of large splits and unbalanced section lengths, we employ an additional layer of text splitting. Based on our experiments and evaluations, we use the 75th percentile of section length as the optimal chunk size to minimize excessively large chunks. The overall pipeline is illustrated in Figure \ref{fig:splitting}.

\subsubsection{Embedding model.} Generating embeddings for documents is a critical aspect of information retrieval. These embeddings act as vector representations of text, enabling text analysis within a vector space. This capability is essential for tasks like semantic search, where the goal is to identify pieces of text with similar meanings. Our evaluation compares the alternatives for embedding models such as SentenceTransformer \cite{reimers-2019-sentence-bert} and BGE \cite{bge_embedding}.

\subsubsection{Indexing and Storage:} One common method for storing and searching over unstructured data involves embedding it and storing the resulting vectors. At query time, we embed the unstructured query and retrieve the most similar embedding vectors. We explored several reputable, open-source vector stores such as ChromaDB\footnote{\url{https://www.trychroma.com}} and FAISS\footnote{\url{https://faiss.ai}}, provided by LangChain\footnote{\url{https://www.langchain.com}}.

\subsection{Query Interpreter}
The system enables interaction with end-users, processing queries as questions and providing answers from a knowledge base. We developed the `Query Interpreter' (UQI) module using natural language processing to interpret queries, extract key terms, and validate information through user confirmation. The module focuses on detecting the following items from the input query: 1) the name of the \textbf{policy} or security rule, 
2) the name of the \textbf{Standard}, and
3) \textbf{subject} (i.e., the network deployment, service, or device).
To discover the optimal language model-prompting combination, we test various large language models (LLMs) like Flan-T5-XXL, Bloom, and Falcon-Instruct-7b, exploring different prompt techniques to optimize accuracy. Our findings indicate that using separate prompts yields the most reliable results with open-source LLMs compared to using one complex prompt seeking all information at once. 

In order to verify the quality and robustness of the generated prompt and also the employed LLM in generating accurate responses, we built a dataset by following the steps below: \textbf{(1) Step 1: Generic Question Templates.} We created sample user question templates with placeholders for various potential values, covering a broad range of queries related to IoT service deployment, compliance, security, and configurations in networked environments. 
Some of the questions target specific policies or security items, requiring direct responses on compliance or configuration. The rest are more general, testing the system’s ability to handle broad information retrieval across relevant policies and regulations. This structure evaluates the system's capability to address both specific and general queries. Then the placeholders are replaced with the actual names of devices, policies, and IoT services. We generate 51 sample Questions each representing one topic. \textbf{(2) Augmented list of Questions.} For each specific question, ChatGPT-4 \footnote{\url{https://openai.com/index/gpt-4/}} was used to produce 10 paraphrased versions, enriching our dataset with diverse linguistic expressions. This enhances vocabulary, sentence structures, and formality levels, ensuring flexibility and adaptability in addressing user queries. \textbf{(3) Manual Data Cleaning and Evaluation.} We meticulously reviewed annotations and labels to ensure dataset accuracy and consistency, updating it when the AI couldn't extract the names of subjects or policies. Consequently, we created an accurate dataset of 191 questions resembling user queries, serving as the initial interaction point between users and the system based on their embeddings.


\subsection{Policy Extraction}
The Policy Definition Extraction module automates the process of linking queries to the standard document corpus and extracting relevant policies for specified security controls. After receiving key terms from the Query Interpreter (QI) module, it identifies the document segments most semantically similar to these terms. The module then compiles a list of relevant policies and security controls with their references to provide a comprehensive response. To optimize performance and inference for each document, we determine the most effective similarity threshold for comparing embeddings. This threshold is also adjusted when new datasets are added to the system.

\subsection{Policy Tagging and Classification}
The Security Policy-Network Configuration Mapping Module translates relevant policies identified by the Policy Definition/Extraction Module into actionable network configuration insights. This module first ensures that existing infrastructure capabilities align with extracted policies and may later advance to verify specific content settings and conduct a thorough risk management analysis.
We utilized LangChain's capabilities alongside a tagging chain, employing a Large Language Model (LLM) to classify text into designated classes. This method leverages LLMs to interpret complex texts and classify them according to the detailed requirements of security policies. Using varied prompts, we extracted critical compliance information for network configurations. Our approaches include: 1) Sentence-based: Analyzing each sentence separately to compile attributes, and 2) Paragraph-based: Reviewing entire paragraphs through a single query.
These techniques require specific post-processing to ensure consistent results.

\subsection{Response Generation}
Finally, the output of the module will be the list of sections and policy controls that are related to the query and should be considered for the downstream tasks such as compliance checks, risk assessment, etc. AuditNet provides tailored, accurate, and actionable answers that directly support the implementation and management of security standards in complex IT infrastructures. 

\section{Conclusion}
In this paper, we have introduced a versatile conversational AI assistant framework called AuditNet, designed to streamline compliance checking across various documents. We demonstrated our prototype using cybersecurity data for our evaluation, however, the proposed framework is adaptable to a wide range of domains, such as legal contracts, educational standards, environmental regulations, and government policies. By automating the processing, indexing, and retrieval of extensive documentation, our framework addresses the challenges of labour-intensive and time-consuming manual methods. It enhances adaptability and scalability, ensuring that professionals can keep up with evolving standards and regulations. This framework aims for efficient and accurate handling of compliance-related tasks, providing tailored, actionable insights to ensure regulatory adherence in diverse professional settings.

\section*{Acknowledgement}
The work has been supported by the Cyber Security Research Centre Limited whose partially funded by the Australian Government’s Cooperative Research Centres Programme. 

\bibliographystyle{ACM-Reference-Format}
\bibliography{reference}

\end{document}